\documentclass[
  a4paper,
  fontsize=11pt,
  ]{article}
\pdfoutput=1

\usepackage{ILD}
\usepackage{graphicx}

\usepackage[utf8]{inputenx}
\usepackage[T1]{fontenc}
\usepackage[british]{babel}
\usepackage{csquotes}

\usepackage{subcaption}
\usepackage{import}
\usepackage{xspace}
\usepackage{graphicx}
\usepackage[shortcuts]{extdash}
\usepackage{siunitx}
\DeclareSIUnit \lightspeed {\text{{c}}}
\usepackage{xcolor}
\definecolor{linkblue}{HTML}{264772}
\usepackage{hyperref}
\hypersetup{
    colorlinks=true,
    linktocpage=true,
    linkcolor=linkblue,
    citecolor=linkblue,
    urlcolor=linkblue
}

\usepackage{wrapfig}

\graphicspath{ {../pictures/} }
\DeclareGraphicsExtensions{.pdf,.png,.jpg}

\newcommand{\dEdx}{\ensuremath{\mathrm{d}E/\mathrm{d}x}\xspace}
\newcommand{\dedx}{\dEdx}

\begin{document}

\hyphenation{
  am-pli-fi-ca-tion
  col-lab-o-ra-tion
  per-for-mance
  sat-u-rat-ed
  se-lect-ed
  spec-i-fied
}

\title{Charged Hadron Identification with dE/dx and Time-of-Flight at Future Higgs Factories}
\date{\today}

\addauthor{Ulrich Einhaus}{\institute{1}\institute{2}}
\addinstitute{1}{Deutsches Elektronen-Synchrotron DESY, Notkestr. 85, 22607 Hamburg, Germany}
\addinstitute{2}{Universität Hamburg, Department of Physics, Jungiusstraße 9, 20355 Hamburg, Germany}
%\emailAdd{ulrich.einhaus@desy.de}
%\note{On behalf of the ILD concept group.}

%\subject{Poster presented at the European Physical Society conference on high energy phsyics (EPS-HEP2021), Hamburg, Germany, July 27th 2021.}
%\publishers{Deutsches Elektronen\-/Synchrotron, DESY}

\abstract{
The design of detector concepts has been driven for a long time by requirements on transverse momentum, impact parameter and jet energy resolutions, as well as hermeticity. Only rather recently it has been realised that the ability to idenfity different types of charged hadrons, in particular kaons and protons, could have important applications at Higgs factories like the International Linear Collider (ILC), ranging from improvements in tracking, vertexing and flavour tagging to measurements requiring strangeness-tagging. While detector concepts with gaseous tracking, like a time projection chamber (TPC), can exploit the specific energy loss, all-silicon-based detectors have to rely on fast timing layers in front of or in the first layers of their electromagnetic calorimeters (ECals). This work will review the different options for realising particle identification (PID) for pions, kaons and protons, introduce recently developed reconstruction algorithms and present full detector simulation prospects for physics applications using the example of the International Large Detector (ILD) concept.
}

\titlecomment{This work was carried out in the framework of the ILD detector
concept group}

\ildproc{phys}{2021}{009}

%\FullConference{ The European Physical Society Conference on High Energy Physics (EPS-HEP2021),\\
%  26-30 July 2021\\
%  Online conference, jointly organized by Universität Hamburg and the research center DESY
%}

\titlepage

%\clearpage
% \flushbottom
% Abstract
% Introduction, PID at ILD
% dE/dx PID
% - method
% - performance
% - application
% TOF PID
% - method
% - performance
% - application
% - improvements
% References

\newpage
\section{Introduction}
\begin{figure}[!hbt]
  \centering
    \includegraphics[width=.85\textwidth,keepaspectratio=true]{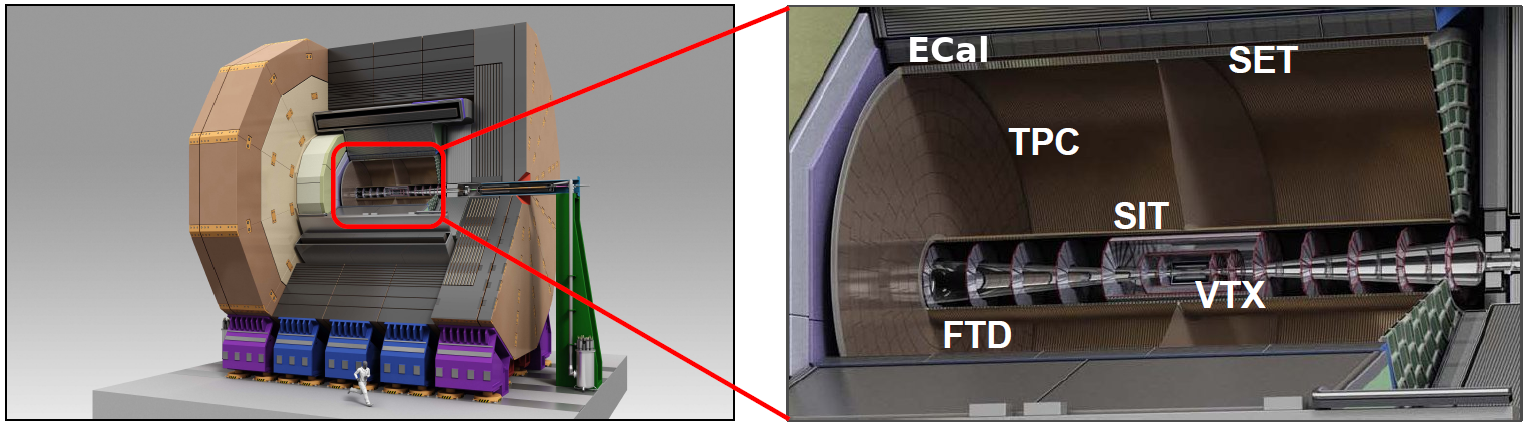}
    \hspace{.1cm}
    \includegraphics[width=.13\textwidth,keepaspectratio=true]{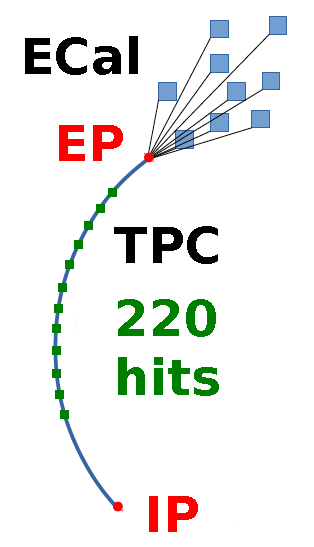}
  \centering
    \caption{Schematic view of ILD, from \cite{ILD_IDR}, and scheme of hits in the TPC and the Ecal used for \dedx and TOF measurements, repectively.}
  \label{fig:ILD}
\end{figure}

The ILC \cite{ILC_TDR_Summary} is a proposed \SI{250}{} - \SI{500}{GeV} $e^+e^-$-collider and ILD \cite{ILD_IDR} is one of its detector concepts, shown in \autoref{fig:ILD}.
It is a multi-purpose detector with a silicon vertex tracker (VTX), a TPC with a silicon envelope (SIT, SET) as central tracking system and a highly granular calorimeter system inside a \SI{3.5}{T} solenoid and a muon system outside of it.
With its forward tracker (FTD) and forward calorimeter system it achieves a high degree of hermeticity.
ILD is designed for particle flow \cite{Pandora} and has an asymptotic momentum resolution of \SI{2e-5}{GeV^{-1}} and a jet energy resolution of better than \SI{3.5}{\percent} above \SI{100}{GeV}.
This work concentrates on the PID capabilities of ILD via measurement of the specific energy loss \dedx in the TPC and via measurement of the time-of-flight (TOF) in the ECal in full-detector simulation \cite{ILD_IDR}.
%The simulation set-up is detailed in \cite{ILD_IDR}.
This work makes use of a large MC production in 2018 \cite{Production_2018}, which generated, simulated and reconstructed about \SI{500}{fb^{-1}} of ILC integrated luminosity.
This includes the entire Standard Model processes and single particles for calibration and detailed studies.
%Each single particle event consists of one charged particle placed at the interaction point (IP) and no background.
%The five species used are electrons, muons, pions, kaons and protons with 100,000 events each and a distribution of the momentum $p$ which is roughly flat in $\log p$.

%\section{Simulation Set-Up}
%
%ILD is simulated using the iLCSoft framework \cite{lcsoft,ilcsoft_desy,ilcsoft_git}.
%The event data format is defined by LCIO \cite{lcio} and the detector geometry description is provided by DD4HEP \cite{dd4hep}.
%Events are generated using Whizard \cite{Whizard,O_Mega}, the event development and detector response are simulated with Geant4 \cite{Geant4_03} and the events are reconstructed using the Marlin package \cite{Marlin}, including Pandora particle flow \cite{Pandora}.

\section{Specific Energy Loss Measurement \dedx}

\begin{figure}
\centering
\begin{minipage}{.52\textwidth}
  \centering
    \includegraphics[width=.95\textwidth,keepaspectratio=true]{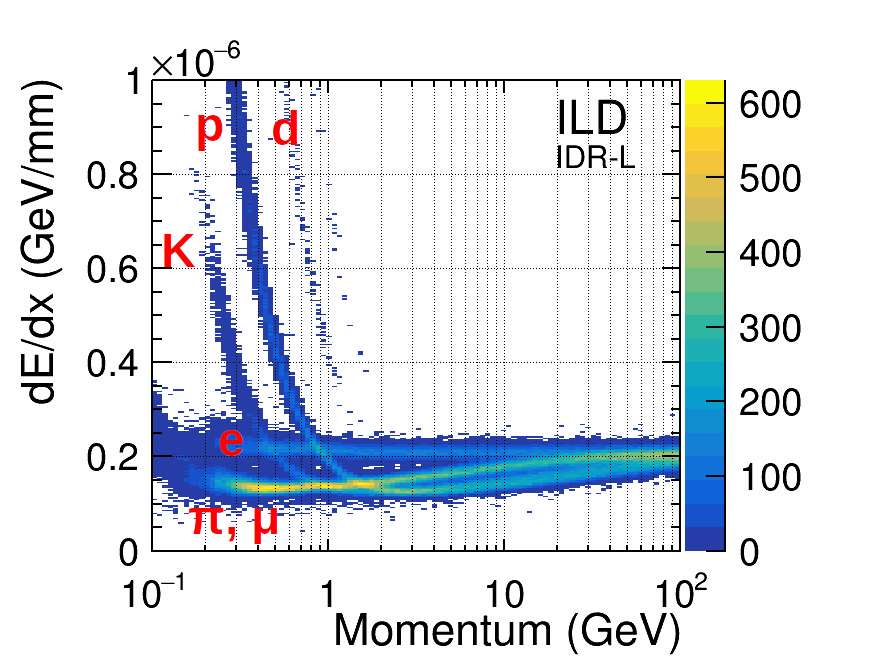}
  \captionof{figure}{Bethe-Bloch curves of single particles. The curves are well separable in most areas, but overlap in some.}
  \label{fig:BB_labels}
\end{minipage}%
\hspace{.2cm}
\begin{minipage}{.45\textwidth}
  \centering
  \includegraphics[width=.9\textwidth,keepaspectratio=true]{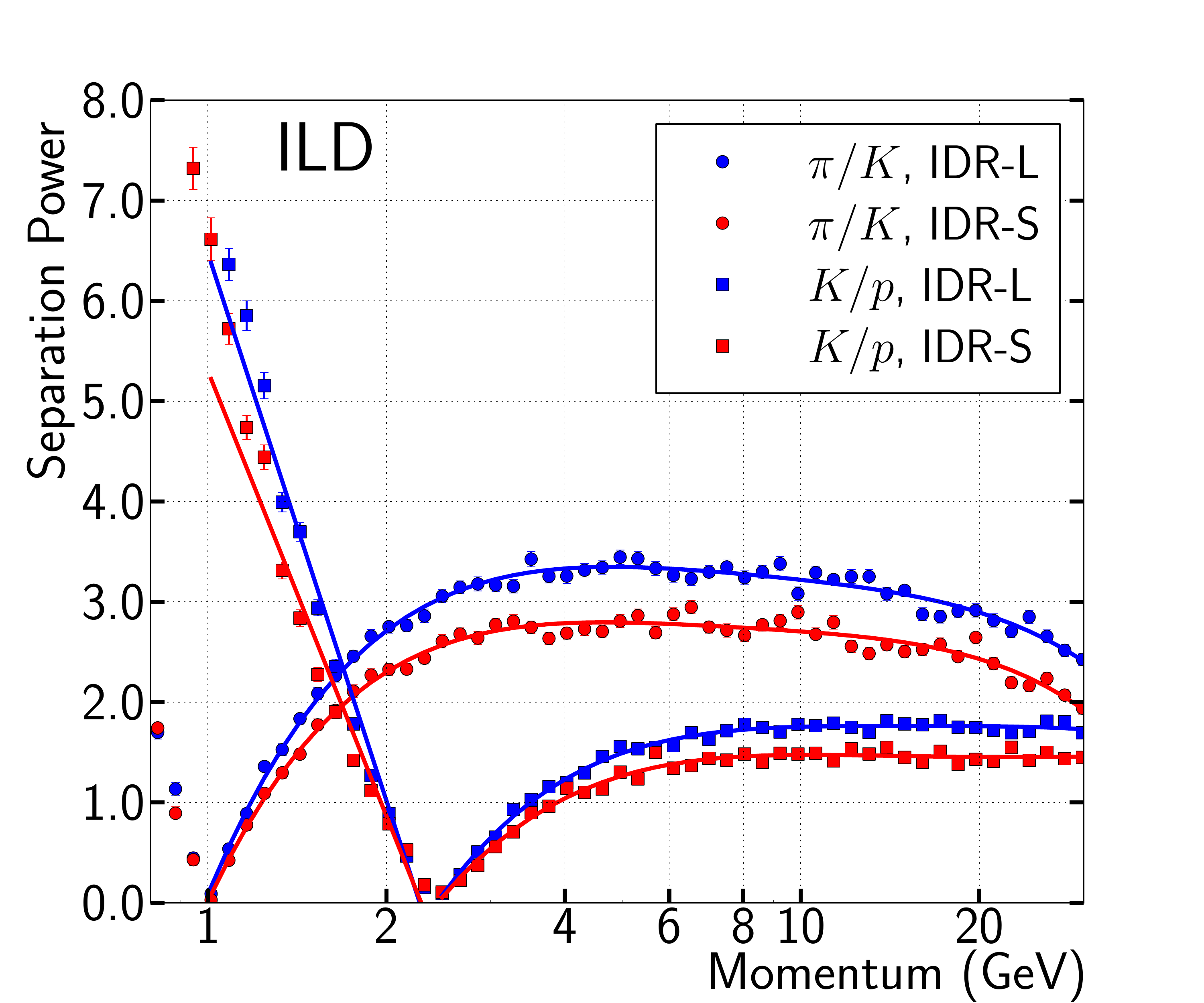}
  \captionof{figure}{Pion-kaon and kaon-proton separation power, for the default and a smaller detector model. The curves are only to guide the eye.}
  \label{fig:SP_ls}
\end{minipage}
\end{figure}

%The ILD TPC is equipped with micro-pattern gaseous detectors (MPGD) as amplification and readout system, and provides up to 220 hits per track.
The ILD TPC provides up to 220 hits per track.
%Its row-based pad readout consists of 220 concentric rows with a height of \SI{6}{mm} covering the TPC's radial extent of \SI{329}{mm} $< r <$ \SI{1770}{mm}, which includes the field cage wall, providing up to 
%The energy depositions of charged particles are reconstructed into one hit per row, so a particle with a transverse momentum larger than about \SI{1}{GeV} that is emitted into the barrel region traverses the entire TPC and creates 220 hits.
%The energy deposition {\ensuremath{\mathrm{d}E}\xspace} of a hit divided by its associated piece of the flight path {\ensuremath{\mathrm{d}x}\xspace} is the \dedx of that hit.
The \dedx estimate for one track is a trimmed truncated mean of the hit \dedx values, rejecting the \SI{8}{\percent} hits with the smallest \dedx values and \SI{30}{\percent} hits with the largest ones.
%The specific energy loss depends on the mass and momentum of an incident particle, which is displayed for single particles in \autoref{fig:BB_labels}.
%The resulting Bethe-Bloch bands of the five simulated species are well identifiable.
The Bethe-Bloch bands of the five simulated single-particle species are displayed in \autoref{fig:BB_labels} and are mostly well identifiable.
%Since Geant4 includes the simulation of hard scattering of the incident particle in the material of the vertex detector, even some deuterium and possibly tritium nuclei can be seen.
The relative \dedx resolution in the simulation is about \SI{4.5}{\percent}, which was adjusted to reflect the \dedx resolution measured in test beam experiments, e.g.\ \cite{JointPaper}, and meets the aim of ILD to have a \dedx resolution of \SI{5}{\percent} or better.
%With a highly granular readout, which is also in development, one could make use of the cluster-counting approach and achieve a \dedx -resolution as good as \SI{3.5}{\percent} \cite{GridPix_2019,Einhaus21}.
The separation power $S$ is the relative distance between the Bethe-Bloch bands, defined as
$S = |\mu_1-\mu_2|/\sqrt{\frac{\sigma_1^2+\sigma_2^2}{2}}$
with $\mu_i$ and $\sigma_i$ being the mean and width of the band of particle $i$, respectively.
\autoref{fig:SP_ls} shows the $\pi/K$ and $K/p$ separation power. %, comparing the two detector models.
%The smaller TPC of the detector model IDR-S translates directly into a lower separation power by about a factor of 1.2.
%$S > 3$ is often referred to as threshold for effective $\pi/K$ separation, which is the case 
$S > 3$ is achieved for particle momenta between about \SI{2}{} and \SI{20}{GeV} in the default detector model IDR-L.
%Notably this threshold marks exactly the difference between the two detector models, with IDR-S staying below that value for the full momentum range.

\begin{figure}
\centering
\begin{minipage}{.48\textwidth}
  \centering
    \includegraphics[width=.8\textwidth,keepaspectratio=true]{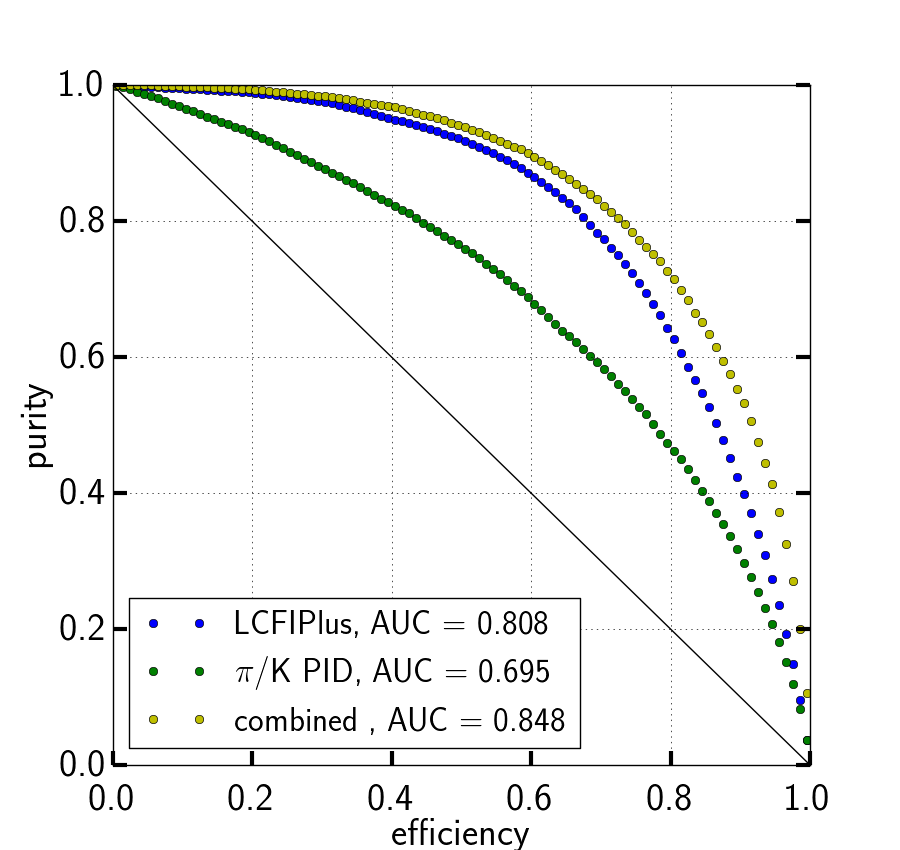}
  \captionof{figure}{BDT result ROC curve for the separation of 1st and 2nd generation hadronic W decays. \dedx -based PID is compared to the vertex-based LCFIPlus flavour tag.}
  \label{fig:ROC}
\end{minipage}%
\hspace{.2cm}
\begin{minipage}{.48\textwidth}
  \centering
  \includegraphics[width=.8\textwidth,keepaspectratio=true]{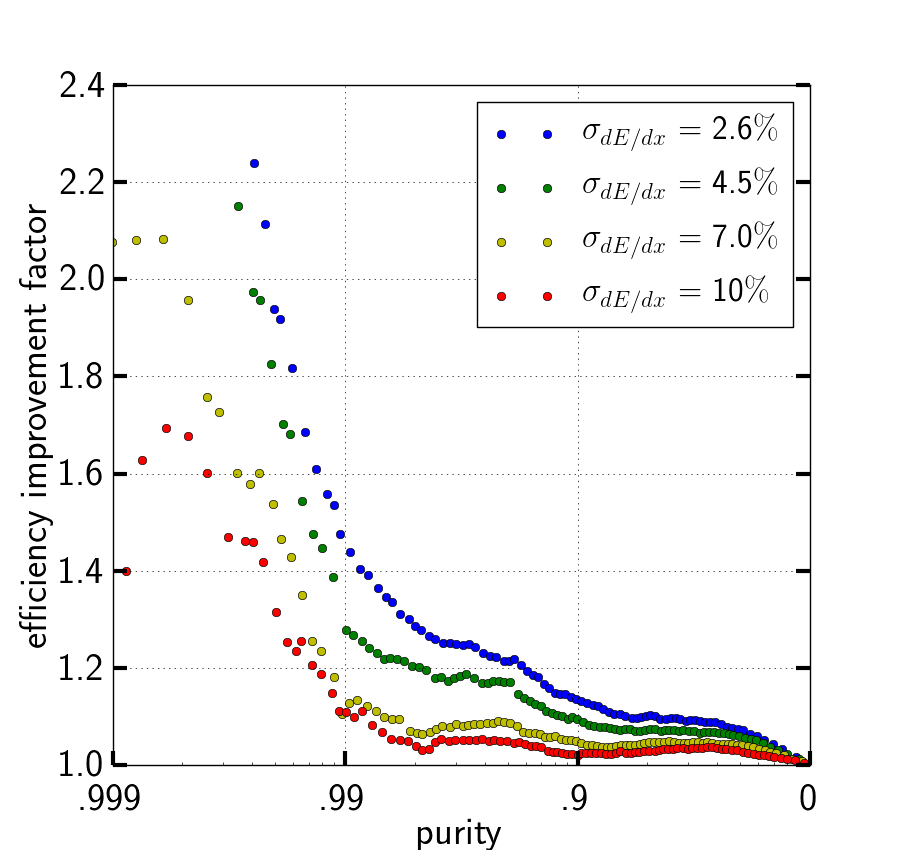}
  \captionof{figure}{Efficiency improvement when adding $\pi$/K PID to the info from LCFIPlus, depending on the requested purity level and on the simulated \dedx resolution.}
  \label{fig:EffImp}
\end{minipage}
\end{figure}

In \cite{Einhaus21}, one example application for PID was investigated, namely its effect on the flavour tag in hadronic W decays.
%The events selected from the MC production were $H \rightarrow WW$ in the semileptonic case. Decays with $\tau$ leptons were rejected, as were cases where the hadronically decaying W decayed cross-generationally.
A boosted decision tree (BDT) was trained to separate decays to a d and a u quark from decays to an s and a c quark.
Since s and c quarks generate more kaons and kaons with higher momenta than d and u quarks, PID can be used in this separation.
The BDT had 20 input variables based on the W-decay system properties, including number, fraction of the jet momentum and leadingness of pions and kaons, which were identified via their \dedx .
These input variables were compared to the case when the default vertex-based flavour tag of ILD, LCFIPlus \cite{lcfiplus16}, was used as BDT input, as well as to a combination of the two.
\autoref{fig:ROC} shows the ROC curves resulting from the BDT training.
The $\pi/K$ PID adds independent information, which increases the area-under-the-curve (AUC) by about \SI{4}{\percent}-points in the combination.
% performs only moderately compared to the LCFIPlus flavour-tag, but it 
This increase translates for a given purity into an improved effiency, i.e.\ in increase in available statistics, 
%In physics measurements, often a high level of flavour purity is required.
%In these cases, adding $\pi/K$ information to the default flavour tag increases the efficiency and thus the available statistics considerably,
which is shown in \autoref{fig:EffImp}.
In addition, the reconstruction and analysis were repeated with different values for the \dedx resolution, demonstrating its impact.
For a required purity between 0.9 and 0.99, the efficiency increase is about \SI{10}{} to \SI{30}{\percent} for the default \dedx resolution.
With an enhanced resolution of \SI{2.6}{\percent}, this would improve to \SI{15}{} to \SI{50}{\percent}, while a reduced resolution of \SI{7}{\percent} or worse would only give an increase of \SI{5}{} to \SI{10}{\percent}.
For very high required purities > 0.99, the efficiency increase can be as large as a factor of 2.
One possible application of this study is the measurement of the central element of the CKM matrix $V_{\mathrm{cs}}$ without assuming unitarity \cite{Vcs_Delphi98}.

\section{Time-of-Flight Measurement TOF}

\begin{figure}
\centering
\begin{minipage}{.5\textwidth}
  \centering
    \includegraphics[width=.96\textwidth,keepaspectratio=true]{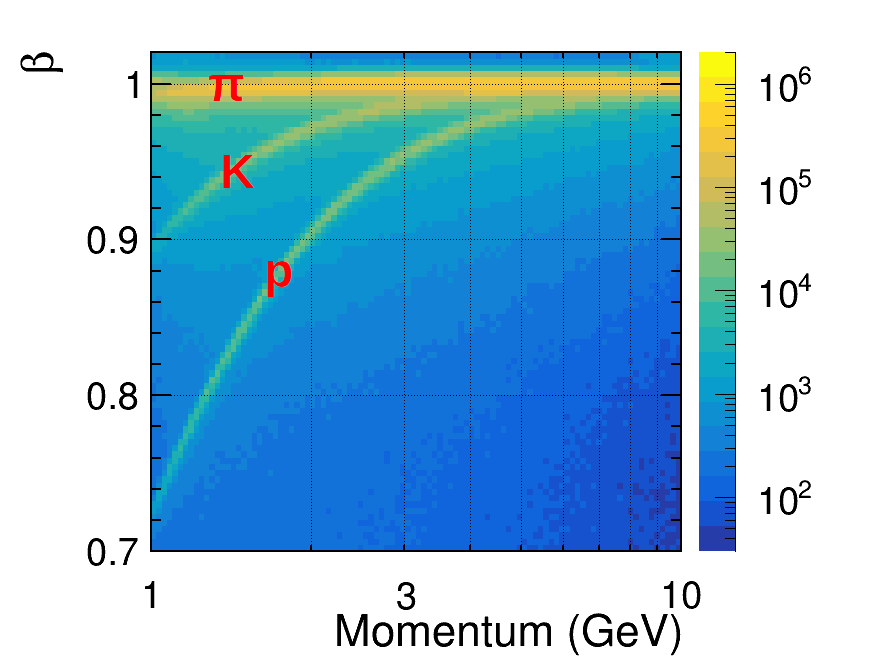}
  \captionof{figure}{TOF $\beta$ curves of particles in full physics events.}
  \label{fig:TOF_beta}
\end{minipage}%
\hspace{.2cm}
\begin{minipage}{.46\textwidth}
  \centering
  \includegraphics[width=.96\textwidth,keepaspectratio=true]{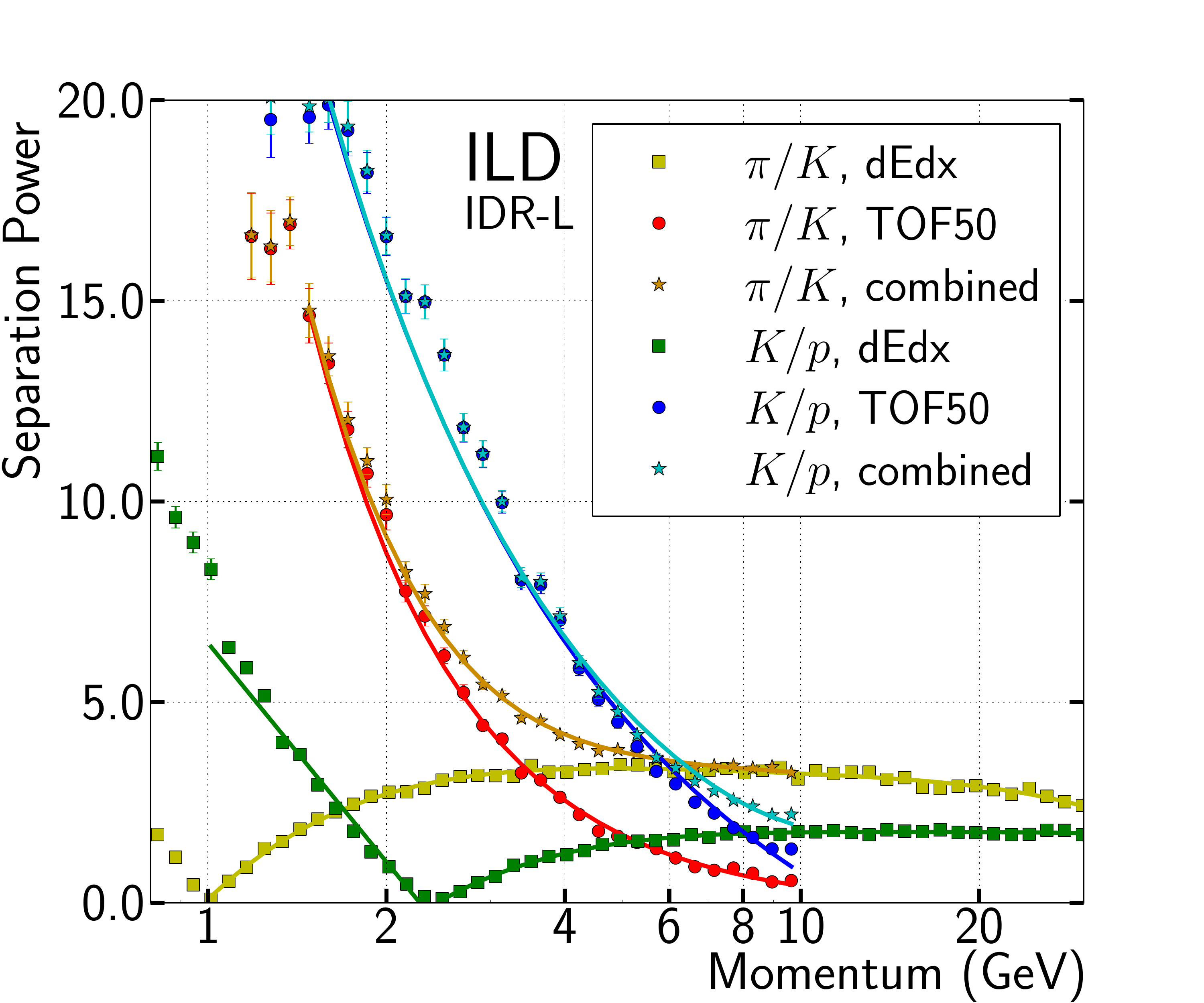}
  \captionof{figure}{Combined \dedx and TOF separation power for pion/kaon and kaon/proton separation. The curves are only to guide the eye.}
  \label{fig:SP_comb}
\end{minipage}
\end{figure}

%While \dedx is a long-established measurement for PID in high-energy particle detectors, improvements in timing resolution have enabled an effective time-of-flight PID rather recently.
For the TOF measurement, a timing resolution of \SI{50}{ps} per channel has been assumed to be achievable for the electromagnetic calorimeter (ECal) and was implemented in the simulation.
%The one hit per layer of the first 10 ECal layers which is closest to the extrapolated track is used to calculate an average as TOF estimator.
The TOF estimator for an incident particle is calculated using the first 10 layers of the ECal.
The timing values of the one active channel in each layer which is closest to the extrapolated track are projected back to the entry point (EP) into the ECal assuming propagation with the speed of light and then averaged.
Together with the track length, the absolute velocity of the particle $\beta$ in units of $c$ is calculated.
Possible improvements on this method are being discussed in \cite{Timing_2021}.
This $\beta$ is shown in \autoref{fig:TOF_beta}, with bands of pions, kaons and protons which are well separable up to \SI{3}{GeV} for $\pi/K$ and \SI{6}{GeV} for $K/p$.
The particles used here are from full physics events, which adds considerable background to the bands.
The separation power, as defined above, can be calculated and combined with the one from \dedx in quadrature, which are both shown in \autoref{fig:SP_comb}.

\section{Conclusions}

Measurements of dE/dx and TOF provide sensitivity for $\pi/K$ and $K/p$ separation in complementary momentum ranges.
In particular TOF excels in the `blind spots' of \dedx where the Bethe-Bloch bands overlap.
%While \dedx extends to high momenta, TOF works very well at low momenta and can cover the `blind spots' of \dedx at \SI{1}{GeV} and \SI{2.5}{GeV} for $\pi/K$ and $K/p$ separation, resp., where the corresponding Bethe-Bloch bands overlap, making \dedx and TOF complementary measurements.
Proven performances from beam tests of technological prototypes have been implemented in detailed full-ILD simulations.
They show promising application possibilities, like enhanced flavour tagging, and open the door for further studies.
This makes PID an invaluable tool to utilize collider data at a future Higgs factory to the best of our possibilities.

%\section{Acknowledgments}

%\newpage
\bibliographystyle{JHEP}
\bibliography{References_reduced}
%\printbibliography
\end{document}